\def\pmb#1{\setbox0=\hbox{#1}%
   \kern-.025em\copy0\kern-\wd0
   \kern.05em\copy0\kern-\wd0
   \kern-0.025em\raise.0433em\box0}
\def\gta{\mathrel{{\lower 3pt\hbox{$\mathchar"218$}}\hskip-8pt
   \raise 2pt\hbox{$\mathchar"13E$}}}
\def\lta{\mathrel{{\lower 3pt\hbox{$\mathchar"218$}}\hskip-8pt
   \raise 2pt\hbox{$\mathchar"13C$}}}
\def\half{{\scriptstyle{1\over2}}}

\def\today{\number\day\space\ifcase\month\or
  January\or February\or March\or April\or May\or June\or
  July\or August\or September\or October\or November\or December\fi
 \space\number\year}
\documentstyle[twoside]{article}

\catcode`\@=11
\long\def\@makefntext#1{
\protect\noindent \hbox to 3.2pt {\hskip-.9pt  
$^{{\eightrm\@thefnmark}}$\hfil}#1\hfill}		

\def\@makefnmark{\hbox to 0pt{$^{\@thefnmark}$\hss}}	
	
\def\ps@myheadings{\let\@mkboth\@gobbletwo
\def\@oddhead{\hbox{}
\rightmark\hfil\eightrm\thepage}   
\def\@oddfoot{}\def\@evenhead{\eightrm\thepage\hfil
\leftmark\hbox{}}\def\@evenfoot{}
\def\sectionmark##1{}\def\subsectionmark##1{}}

\oddsidemargin=\evensidemargin
\newcounter{sectionc}\newcounter{subsectionc}\newcounter{subsubsectionc}
\renewcommand{\section}[1] {\vspace{12pt}\addtocounter{sectionc}{1} 
\setcounter{subsectionc}{0}\setcounter{subsubsectionc}{0}\noindent 
	{\tenbf\thesectionc. #1}\par\vspace{5pt}}
\renewcommand{\subsection}[1] {\vspace{12pt}\addtocounter{subsectionc}{1} 
	\setcounter{subsubsectionc}{0}\noindent 
	{\bf\thesectionc.\thesubsectionc. {\kern1pt \bfit #1}}\par\vspace{5pt}}
\renewcommand{\subsubsection}[1] {\vspace{12pt}\addtocounter{subsubsectionc}{1}
	\noindent{\tenrm\thesectionc.\thesubsectionc.\thesubsubsectionc.
	{\kern1pt \tenit #1}}\par\vspace{5pt}}
\newcommand{\nonumsection}[1] {\vspace{12pt}\noindent{\tenbf #1}
	\par\vspace{5pt}}

\newcounter{appendixc}
\newcounter{subappendixc}[appendixc]
\newcounter{subsubappendixc}[subappendixc]
\renewcommand{\thesubappendixc}{\Alph{appendixc}.\arabic{subappendixc}}
\renewcommand{\thesubsubappendixc}
	{\Alph{appendixc}.\arabic{subappendixc}.\arabic{subsubappendixc}}

\renewcommand{\appendix}[1] {\vspace{12pt}
        \refstepcounter{appendixc}
        \setcounter{figure}{0}
        \setcounter{table}{0}
        \setcounter{lemma}{0}
        \setcounter{theorem}{0}
        \setcounter{corollary}{0}
        \setcounter{definition}{0}
        \setcounter{equation}{0}
        \renewcommand{\thefigure}{\Alph{appendixc}.\arabic{figure}}
        \renewcommand{\thetable}{\Alph{appendixc}.\arabic{table}}
        \renewcommand{\theappendixc}{\Alph{appendixc}}
        \renewcommand{\thelemma}{\Alph{appendixc}.\arabic{lemma}}
        \renewcommand{\thetheorem}{\Alph{appendixc}.\arabic{theorem}}
        \renewcommand{\thedefinition}{\Alph{appendixc}.\arabic{definition}}
        \renewcommand{\thecorollary}{\Alph{appendixc}.\arabic{corollary}}
        \renewcommand{\theequation}{\Alph{appendixc}.\arabic{equation}}
        \noindent{\tenbf Appendix \theappendixc #1}\par\vspace{5pt}}
\newcommand{\subappendix}[1] {\vspace{12pt}
        \refstepcounter{subappendixc}
        \noindent{\bf Appendix \thesubappendixc. {\kern1pt \bfit #1}}
	\par\vspace{5pt}}
\newcommand{\subsubappendix}[1] {\vspace{12pt}
        \refstepcounter{subsubappendixc}
        \noindent{\rm Appendix \thesubsubappendixc. {\kern1pt \tenit #1}}
	\par\vspace{5pt}}

\newcommand{\textlineskip}{\baselineskip=13pt}
\newcommand{\smalllineskip}{\baselineskip=10pt}

\def\eightcirc{
\begin{picture}(0,0)
\put(4.4,1.8){\circle{6.5}}
\end{picture}}
\def\eightcopyright{\eightcirc\kern2.7pt\hbox{\eightrm c}} 

\newcommand{\copyrightheading}[1]
	{\vspace*{-2.5cm}\smalllineskip{\flushleft
	{\footnotesize International Journal of Modern Physics B, #1}\\
	{\footnotesize $\eightcopyright$\, World Scientific Publishing
	 Company}\\
	 }}

\newcommand{\publisher}[2]{{\begin{center}\footnotesize\smalllineskip 
	Received #1\\
	Revised #2
	\end{center}
	}}

\def\abstracts#1#2#3{{
	\centering{\begin{minipage}{4.5in}\baselineskip=10pt\footnotesize
	\parindent=0pt #1\par 
	\parindent=15pt #2\par
	\parindent=15pt #3
	\end{minipage}}\par}} 

\renewenvironment{thebibliography}[1]			
	{\frenchspacing
	 \ninerm\baselineskip=11pt
	 \begin{list}{\arabic{enumi}.}
	{\usecounter{enumi}\setlength{\parsep}{0pt}
	 \setlength{\leftmargin 12.7pt}{\rightmargin 0pt} 
	 \setlength{\itemsep}{0pt} \settowidth
	{\labelwidth}{#1.}\sloppy}}{\end{list}}

\newcounter{itemlistc}
\newcounter{romanlistc}
\newcounter{alphlistc}
\newcounter{arabiclistc}

\newcommand{\fcaption}[1]{
        \refstepcounter{figure}
        \setbox\@tempboxa = \hbox{\footnotesize Fig.~\thefigure. #1}
        \ifdim \wd\@tempboxa > 5in
           {\begin{center}
        \parbox{5in}{\footnotesize\smalllineskip Fig.~\thefigure. #1}
            \end{center}}
        \else
             {\begin{center}
             {\footnotesize Fig.~\thefigure. #1}
              \end{center}}
        \fi}

\newcommand{\tcaption}[1]{
        \refstepcounter{table}
        \setbox\@tempboxa = \hbox{\footnotesize Table~\thetable. #1}
        \ifdim \wd\@tempboxa > 5in
           {\begin{center}
        \parbox{5in}{\footnotesize\smalllineskip Table~\thetable. #1}
            \end{center}}
        \else
             {\begin{center}
             {\footnotesize Table~\thetable. #1}
              \end{center}}
        \fi}

\def\@citex[#1]#2{\if@filesw\immediate\write\@auxout
	{\string\citation{#2}}\fi
\def\@citea{}\@cite{\@for\@citeb:=#2\do
	{\@citea\def\@citea{,}\@ifundefined
	{b@\@citeb}{{\bf ?}\@warning
	{Citation `\@citeb' on page \thepage \space undefined}}
	{\csname b@\@citeb\endcsname}}}{#1}}

\newif\if@cghi
\def\cite{\@cghitrue\@ifnextchar [{\@tempswatrue
	\@citex}{\@tempswafalse\@citex[]}}
\def\citelow{\@cghifalse\@ifnextchar [{\@tempswatrue
	\@citex}{\@tempswafalse\@citex[]}}
\def\@cite#1#2{{$\null^{#1}$\if@tempswa\typeout
	{IJCGA warning: optional citation argument 
	ignored: `#2'} \fi}}

\def\pmb#1{\setbox0=\hbox{#1}
	\kern-.025em\copy0\kern-\wd0
	\kern.05em\copy0\kern-\wd0
	\kern-.025em\raise.0433em\box0}

\def\fnt#1#2{\footnotetext{\kern-.3em
	{$^{\mbox{\scriptsize #1}}$}{#2}}}

\def\fpage#1{\begingroup
\voffset=.3in
\thispagestyle{empty}\begin{table}[b]\centerline{\footnotesize #1}
	\end{table}\endgroup}

\def\runninghead#1#2{\pagestyle{myheadings}
\markboth{{\protect\footnotesize\it{\quad #1}}\hfill}
{\hfill{\protect\footnotesize\it{#2\quad}}}}
\font\tenrm=cmr10
\font\tenit=cmti10 
\font\tenbf=cmbx10
\font\bfit=cmbxti10 at 10pt
\font\ninerm=cmr9

\font\eightrm=cmr8

\def\qed{\hbox{${\vcenter{\vbox{			
   \hrule height 0.4pt\hbox{\vrule width 0.4pt height 6pt
   \kern5pt\vrule width 0.4pt}\hrule height 0.4pt}}}$}}

\def\bsc{{\sc a\kern-6.4pt\sc a\kern-6.4pt\sc a}}	
\def\bflatex{\bf L\kern-.30em\raise.3ex\hbox{\bsc}\kern-.14em 
T\kern-.1667em\lower.7ex\hbox{E}\kern-.125em X} 

\begin{document}

\runninghead{Stripes, Carriers, Pseudogap, and Superconductivity in the
Cuprates} {Stripes, Carriers, Pseudogap, and Superconductivity in the
Cuprates} 

\normalsize\textlineskip
\thispagestyle{empty}
\setcounter{page}{1}

\copyrightheading{}			

\vspace*{0.88truein}

\fpage{1}
\centerline{\bf STRIPES, CARRIERS, PSEUDOGAP, AND}
\vspace*{0.035truein}
\centerline{\bf SUPERCONDUCTIVITY IN THE CUPRATES}
\vspace*{0.37truein}
\centerline{\footnotesize JOSEF ASHKENAZI}
\vspace*{0.015truein} 
\centerline{\footnotesize\it Physics Department, University of Miami,
P.O. Box 248046} 
\baselineskip=10pt
\centerline{\footnotesize\it Coral Gables, FL 33124, U.S.A.}
\vspace*{0.225truein}
\publisher{(June 9, 1999)}{(June 9, 1999)}

\vspace*{0.21truein}
\abstracts{The electronic structure of the high-$T_c$ cuprates is
studied on the basis of both ``large-$U$'' and ``small-$U$'' orbitals. A
striped structure is obtained, and three types of carriers: polaron-like
``stripons'' carrying charge, ``quasielectrons'' carrying charge and
spin, and ``svivons'' carrying spin and lattice distortion. Anomalous
properties of the cuprates and specifically their transport properties
are derived. Pairing is found to result from transitions between pair
states of quasielectrons and stripons through the exchange of svivons.
The pairing results in superconductivity when the stripons conduction is
coherent, and in a pseudogap phase when it is not.}{}{} 


\textlineskip			
\vspace*{12pt}			


\noindent 
Both ``large-$U$'' and ``small-$U$'' orbitals are considered in the
vicinity of the Fermi level ($E_{_{\rm F}}$) of the high-$T_c$
cuprates\cite{Ashk}. The large-$U$ orbitals are treated using the
``slave-fermion'' method\cite{Barnes}, where the electrons are described
in terms of auxiliary particles: fermion "holons" and  boson "spinons".
An extended auxiliary Hilbert space is introduced, and physical
observables are expressed in terms of the auxiliary-space Green's
functions. 

Bose condensation of the spinons results in antiferromagnetism (AF). It
has been shown\cite{Emery} that a lightly doped AF plane segregates into
AF stripes and narrower ``charged'' stripes forming antiphase domain
walls between them. Such a scenario is supported by
experiment\cite{Bian,stripes}, and there exists growing evidence
that it probably exists, at least dynamically on the short range, in all
the superconducting cuprates. 

Spin-charge separation applies along the one-dimensional charged stripes
(where such an approximation is known to be valid), and holons within
them are referred to as ``stripons'', carrying charge, but not spin.
Since\cite{stripes} the stripes in the cuprates are disordered, and
consist of disconnected segments, it is assumed here that an appropriate
starting point is of localized stripon states. 

Other carriers (of both charge and spin) result from the hybridization
of small-$U$ states and coupled holon-spinon states which are orthogonal
to the stripon states. These carriers as referred to as
``Quasi-electrons'' (QE's), and their bare energies form
quasi-continuous ranges of bands crossing $E_{_{\rm F}}$. 

These fields are coupled to each other due to hopping and hybridization
terms of the orbitals, and the coupling can be expressed in terms of an
effective Hamiltonian\cite{Ashk}. It introduces a vertex between the QE,
stripon, and spinon propagators, and ``vertex corrections'' are
negligible by a generalized Migdal theorem. For sufficiently doped
cuprates the self-consistent self-energy corrections determine
quasiparticles of the following features: 

The spinon spectral functions are proportional to $\omega$ for small
$\omega$, resulting in the absence of long-range AF order (though
short-range order persists). The energies of the localized stripon
states are renormalized to a very narrow range around zero, thus getting
polaron-like states. Some hopping via QE-spinon states results is the
onset of coherent itineracy at low temperatures, with a bandwidth of
$\sim$$0.02\;$eV. The stripon scattering rates scale as: $\Gamma^p({\bf
k}, \omega) \propto A \omega^2 + B \omega T + CT^2$. The QE scattering
rates scale as: $\Gamma^q({\bf k}, \omega)\propto T$ for $T\gg
|\omega|$, and $\Gamma^q({\bf k}, \omega)\propto\half |\omega|$ for
$T\ll |\omega|$, in agreement with ``marginal Fermi liquid''
phenomenology\cite{Varma}. 

It was found\cite{Bian} that the charged stripes are characterized by
LTT structure, while the AF stripes are characterized by LTO structure.
The result would be that spinons are ``dressed'' by phonons when they
are emitted or absorbed in processes where a stripon is transformed into
a QE, or vice versa.  Such a phonon-dressed spinon is referred to as a
``svivon'', and it carries spin and lattice distortion. 

The optical conductivity of the doped cuprates has two
components\cite{Tanner}, a Drude term and mid-IR peaks. The Drude term
is due to transitions between QE states, while the mid-IR peaks are due
to excitations of stripon states. The electronic spectral function,
measured in photoemission experiments, has a ``coherent'' part, due to
the contributions of few QE bands, and an ``incoherent'' part of a
comparable weight, due to the contributions of other quasi-continuous QE
bands, and stripon-svivon states. The observed ``Shadow bands'' and
``extended'' van Hove singularities result from the effect of the
striped superstructure on the QE bands\cite{Salk}. 

The electric current is expressed as a sum ${\bf j} = {\bf j}^q + {\bf
j}^p$ of QE and stripon contributions. Since stripons transport occurs
through transitions to intermediate QE-svivon states, one gets ${\bf
j}^p \cong \alpha {\bf j}^q$, where $\alpha$ is approximately
$T$-independent. In order for this condition to be satisfied gradients
of the QE and stripon chemical potentials must be formed in the presence
of an electric field or a temperature gradient. 

It has been shown\cite{Ashk} that the temperature dependence of the
electrical resistivity can then be expressed as $\rho = (D+CT+A+BT^2)/
N$, and of the Hall constant as $R_{_{\rm H}} = \rho /
\cot{\theta_{_{\rm H}}}$, where $(\cot{\theta_{_{\rm H}}})^{-1} =
Z(D+CT)^{-1} + (A+BT^2)^{-1}$. The $(D+CT)$ and $(A+BT^2)$ terms are
(respectively) due to the QE and stripon scattering rates discussed
above (to which temperature-independent impurity scattering terms are
added). These expressions reproduce\cite{Ashk} the systematic behavior
of the transport quantities in different cuprates\cite{Chien}. 

It has also been shown\cite{Ashk} that the thermoelectric power (TEP)
$S$ can be expressed in terms of QE and stripon terms as $S=(N^qS^q +
N^pS^p)/(N^q + N^p)$, where $S^q\propto T$, while the stripon term
saturates at $T \simeq 200\; $K to $S^p=(k_{_{\rm B}}/{\rm
e})\ln{[(1}-n^p)/n^p]$. This result is consistent with the typical
behavior of the TEP in the cuprates\cite{TEP}. 

The present approach\cite{Ashk} provides a pairing mechanism involving
transitions between pair states of QE's and stripons through the
exchange of svivons. This is conceptually similar to the interband pair
transition mechanism proposed by Kondo\cite{Kondo}. A condition for
superconductivity is that the narrow stripon band maintains coherence
between different stripe segments. Pairing can, however, occur also when
the stripons are incoherent, and the condensate is then interpreted as
the pseudogap phase found in underdoped cuprates. 

Thus a normal-state pseudogap is expected to have a similar size and
symmetry to that of the superconducting gap, as has been observed, and
its opening accounts for most of the pair-condensation energy, as has
been observed too. In overdoped cuprates, where pairing occurs when the  
stripon states are coherent, a BCS-like behavior of the gap is expected, 
as has been observed\cite{mihail}. 

Stripon coherence is energetically favorable at temperatures where there 
is a clear distinction between occupied and unoccupied stripon band
states. Thus, an estimate for the coherence temperature for an almost
empty (full) stripon band is given by the distance ${\cal E}_{_{\rm F}}$
of the Fermi level from the bottom (top) of the band at $T=0$. This
result agrees with the ``Uemura plots''\cite{Uemura}. The
``boomerang-type'' behavior of the Uemura plots in overdoped
cuprates\cite{Nieder} is consistent with as a transition from a stripon
band top to a band bottom with BCS-type behavior. 

\nonumsection{References}

\end{document}